\documentclass{book}    
\usepackage{piers}  
\begin{document}

\title{
Towards R-Space Bose-Einstein Condensation of Photonic Crystal
Exciton Polaritons} \maketitle

\author      {D. L. Boiko}
\affiliation {Sowoon Technologies S.\'a.r.l.,}
\address     {}
\city        {Lausanne}
\postalcode  {}
\country     {Switzerland}
\email       {dmitri.boiko@epfl.ch}  
\misc        { }  
\nomakeauthor
%
%
\begin{authors}
{\bf D. L. Boiko} 
\\
\medskip
Sowoon Technologies S.\'a.r.l.,
1015, Lausanne, Switzerland
\end{authors}
%
\begin{paper}
\begin{piersabstract}
Coupled states of semiconductor quantum well (QW) excitons and
photons in a two dimensional (2D) periodic lattice of 
microcavities 
are analyzed theoretically, revealing
allowed 
bands and forbidden 
gaps in the energy 
spectrum of
exciton polaritons. Photonic crystal exciton polaritons 
have spatially uniform excitonic constituent set 
by 
flat 
QWs, 
but exhibit periodic Bloch oscillations in the plane of QWs due
to their photonic component.
The envelope functions of 
photonic crystal exciton polaritons can be 
tailored via 
effective potential of 
a photonic crystal heterostructure, by using 
quasi-periodic 
lattices of microcavities. 
Confined envelope function states of lower and upper polaritons
and the Bose-Einstein condensation of lower polaritons are
analyzed here in a photonic crystal heterostructure trap with
harmonic oscillator potential. This concept is numerically
illustrated on example of CdTe/CdMgTe microcavities.

\end{piersabstract}


\psection{Introduction}
Recently, 
several claims have been made on achieving the Bose-Einstein
condensation (BEC) in solids \cite{Kasprzak06, Deng06,
Christopoulos07,Balili07}.
In these 
experiments, 
semiconductor microcavities 
incorporating heterostructure quantum wells (QWs)
sandwiched 
between two distributed Bragg reflectors (DBRs) are used in a
strong-coupling regime, such that 
the coupled states of
QW excitons and 
cavity photons
represent 
composite (bosonic) quasiparticles. Due to photonic constituent,
the cavity exciton polaritons
are of 
light effective masses ($10^{-4}$ of free electron
mass) and
allow 
the macroscopic
quantum degeneracy 
to be achieved at lower density and higher temperature compared to
the BEC transition in an atomic vapor.
Nevertheless, the recent reports on 
k-space BEC and early observations of macroscopic coherence in
polariton system (\textit{e.g.}, Refs.\cite{Deng02,Weihs03}) 
allow one to
argue that these observations may not be attributed uniquely 
to the BEC phase transition 
(\emph{e.g.},
Ref.\cite{Bloch08}).

Incompleteness of experimental data and a short lifetime of the
cavity exciton polaritons 
do not allow a
thermalization 
and spontaneous transition to a macroscopically ordered state to
be unambiguously confirmed. Thus in
\cite{Kasprzak06,Christopoulos07,Balili07}, a
correlation function of the first order $g^{(1)}$ is measured 
to prove 
the fact of macroscopic quantum coherence. 
However,
such
phase correlations may not be attributed uniquely 
to a quantum coherent state, which assumes that $g^{(n)}\equiv 1$
for any order $n$ \cite{Glauber07}. In particular, the first
order correlations can also be observed in a chaotic thermal state
\cite{HBT56,Glauber63A,Glauber63B}.
Therefore, other tests verifying the nature of 
a macroscopically ordered 
state of exciton polaritons are
important.

A harmonic oscillator trap with confining potential $U{=}\frac12
\alpha_{LP} r^2$ for lower polaritons (LPs) can provide an
evidence of BEC by displaying distinct spatial distributions of
polaritons 
in the condensate and non-condensate fractions \cite{Snoke03}. In
\cite{Balili07}, a claim is made on achieving r-space BEC of GaAs
cavity exciton polaritons in a trap produced via excitonic
component 
of polaritons, by introducing the Pikus-Bir deformation potential
$U{=} \frac 12
\alpha_{X} r^2$ for excitons (X) in 
GaAs QWs.
The reported force constant of the trap for lower polaritons 
$\alpha_{LP}{=}480~eV/cm^2$ assumes that the corresponding
trapping potential for excitons is of $\alpha_{X}{=}1150~eV/cm^2$,
which improves by a factor of 30 the force constant of a trap 
previously reported 
for GaAs QW excitons \cite{Voros06}.
However, the 
features 
of the coherent fraction of LPs, which was delocalized over a
region of $8~\mu m$ size, were different from the expected
point-like localization of r-space BEC condensate fraction.
This discrepancy might be attributed to a negative photon-exciton
energy detuning inherent to stress-induced traps, since such
negative energy detuning
prevents thermalization 
of lower polaritons \cite{Deng06}. These allow one to
question 
whether the reported observations indeed might be attributed to
the r-space BEC of cavity exciton polaritons and whether such
traps with negative photon-exciton relative energy are suitable
for BEC experiments. 

In this paper, a 
novel concept of harmonic oscillator trap for 
exciton
polaritons is proposed, 
benefiting from the light propagation features in 
quasi-periodic 
2D arrays 
of optically coupled microcavities. Such 
photonic crystal
lattices 
oriented 
in the plane of QWs
are shown here to induce 
periodic Bloch oscillations in exciton polariton wave functions.
By introducing tailored variations of the cavity pixels 
across an array, 
an effective potential can be superimposed on these oscillations
to control the envelop wave functions of photonic crystal exciton
polaritons. This opens the way to trap polaritons in a harmonic
oscillator potential
at a positive photon-exciton 
detuning, by shaping 
the photonic constituent of polaritons.  Such 
traps 
will favor thermalization of lower polaritons in experiments on
r-space BEC. This concept is numerically illustrated here on
example of harmonic oscillator trap implemented with array of
CdTe/CdMgTe microcavities.

The paper is organized as follows. Sec.~\ref{SecStructure}
details a structure of arrays of coupled microcavities treated
here. In Sec.~\ref{SecUniLatt}, the photonic crystal exciton
polaritons are analyzed in a
uniform-lattice arrays. 
In Sec.~\ref{SecHarmTrap}, a photonic crystal heterostructure
trap for exciton polariton is considered. The properties of
non-condensate and condensate fractions of lower polaritons in
the trap 
are discussed in
Sec.~\ref{SecResDisc}.

\psection{Quasi-periodic array of coupled microcavities}
\label{SecStructure}

\begin {figure}[tbp]
\center
\includegraphics {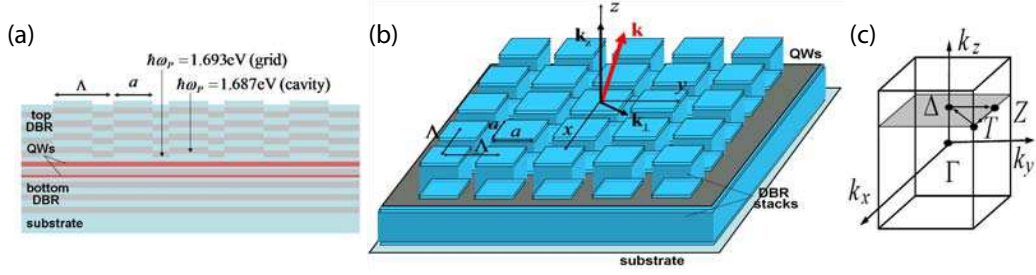}
\caption {Photonic crystal 
harmonic oscillator trap implemented with 
quasi-periodic 2D array of coupled microcavities. (a): Schematic
of the wafer structure with etched array of shallow mesa pixels at
the cavity spacer layer and regrown top DBR structure.
(b): Schematic of the array with parabolically varying 
size of microcavity pixels used to define effective potential for
exciton polaritons. (c): Brillouin zone of a square lattice array
of microcavities. $\Lambda$ is the lattice pitch, $a$ is the
width of the cavity pixels. } \label {fig1}
\end {figure}

The 
microcavities discussed here are arranged in a quasi-periodic two
dimensional lattice (Fig.~\ref{fig1}). Such arrays of optically
coupled microcavities belong to a particular class of paraxial
photonic crystal (PhC) structures, in which the light propagates
mostly normal to the periodic lattice plane
\cite{Boiko07,Boiko07B}. These structures employ lattices of
periods significantly exceeding the optical wavelength. For
example, the arrays of CdTe/CdMgTe microcavities treated here
employ lattices of $3~\mu m$ pitch (Fig.~\ref{fig1}) and have
a vertical cavity structure %
(one-wavelength cavity 
incorporating QWs and sandwiched between two DBRs)
optimized 
at 730 nm wavelength. 
In such structures, only a small transversal component of wave
vector $\mathbf{k}$ of a photon undergoes periodic Bragg
reflections in the optical lattice plane. The main
$\mathbf{k}$-vector component 
(along the cavity $z$-axis in Fig.~\ref{fig1})
is fixed by the cavity 
roundtrip self-repetition condition. 

These arrays 
can be fabricated 
by introducing intermediate processing steps 
during a wafer growth (\emph{e.g.}, shallow mesa-etching) 
 such that 
microcavities 
share multiple QWs 
in 
the $\lambda$-cavity 
and in the few first periods of the bottom DBR. In this way, a
periodic photonic crystal lattice can be defined in the
plane of 
QWs, as indicated in Fig.~\ref{fig1}.
By analogy with the cavity exciton polaritons in 
broad-area microcavities, the coupled states of QW excitons and
photons in coupled arrays of microcavities are
termed here as  
photonic crystal exciton polaritons. (As shown in
Sec.~\ref{SecUniLatt}, their wave functions exhibit periodic
Bloch oscillations in the plane of QWs.)
As 
in the solitary microcavities utilizes 
in experiments on k-space BEC \cite{Kasprzak06, Deng06,
Christopoulos07,Balili07}, there are two degrees of freedom
available for photonic crystal exciton polaritons
to form spontaneously  
a microscopically ordered 
state.

Lattices of CdTe/CdMgTe microcavities analyzed here theoretically
are 
defined by etching a periodic pattern 
of shallow mesa structures at the cavity spacer layer and
subsequently regrowing a complete top DBR structure
(Fig.\ref{fig1}). 
This periodic 
pattern 
is used to define the position of
cavity pixels. 
In the model calculations, 
a vertical composition of the cavity wafer 
is similar to the one of Ref.\cite{Kasprzak06}. In particular, the
cavities incorporating 16 quantum wells with 
exciton energy
of
1.682 eV and Rabi 
energy splitting 
of 26 meV are analyzed. The vertical-cavity modes oscillate at
photon energies of 1.687 and 1.693 eV  at the
cavity pixels 
and 
array grid separating the pixels, respectively. Such tiny
variations in the resonant energy of vertical-cavity modes ($\sim
0.3\%$)
suffice to define a 
lattice of paraxial photonic crystal \cite{Boiko07}. 
As shown below,  low-contrast lattices considered here allow a
forbidden energy gap to be opened in the spectrum
of 
coupled optical modes of entire array structure. Another
important parameter, which impacts the photonic band structure,
is the lattice cell fill factor (FF) defined as the
area ratio of the 
cavity pixel and of the lattice cell. For square arrays treated
here, the lattice period $\Lambda$ is $3~\mu m$ and the lattice
cell fill factor $FF$ varies in the range of $0.5{-}1$
($FF{=}a^2/\Lambda^2$, with $a$ being the square pixel width).

\psection{Photonic crystal exciton polariton}\label{SecUniLatt}

\begin {figure}[tbp]
\center
\includegraphics {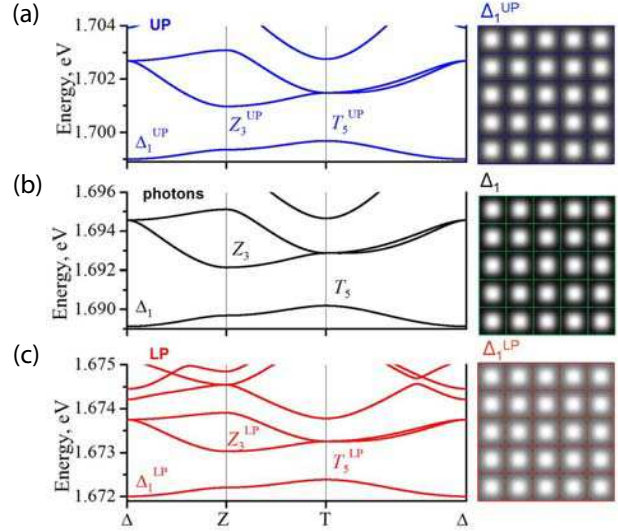}
\caption {Energy bands (left panels) and lowest-energy state wave
functions
$|\psi(x,y)|^2$ (right panels) 
of the upper polaritons (a), photons (b) and lower polaritons (c)
in a periodic lattice of coupled CdTe/CdMgTe microcavities
($FF=0.5$, $\Lambda=3~\mu m$). The panels (a)-(c) are ordered
according the energy scale. }\label {fig2}
\end {figure}

Fig. \ref{fig2} (b) shows optical mode dispersion curves for a
uniform
array of microcavities 
with 
the lattice fill factor of $FF{=}0.5$ ($2.1~\mu m$ 
cavity pixels arranged in a square lattice of $3~\mu m$ pitch).
The photonic
band structure was calculated using the 
paraxial Hamiltonian approach developed in
Refs.~\cite{Boiko07,Boiko07B}. The energy bands are plotted along
the high symmetry lines of the Brillouin zone
[Fig.~\ref{fig1}(c)]. The wave vector $\mathbf k$ of a photon is
$(0,0,k_z)$, $(0,\pi/\Lambda,k_z)$ and
$(\pi/\Lambda,\pi/\Lambda,k_z)$ at the $\Delta$, $Z$ and $T$
points of the Brillouin zone (BZ), respectively.

It can be seen that 
for 
low-contrast 
structures considered here, a complete 2D band gap is opened
between the states in the T and Z points of the BZ
\cite{Boiko07,Boiko07B}. In Fig.
\ref{fig2} (b), the forbidden energy 
gap is of 2 meV. Note that the optical modes originating from
equivalent T points of the BZ show a $\pi$ phase shift between
adjacent lattice sites (the out-of-phase modes) while the modes
located at the Z points of the BZ exhibit the out-of-phase
oscillations along only one lattice direction and they oscillate
in-phase at the lattice sites located along the second direction
of the lattice \cite{Boiko07,Boiko07B}.

Within the framework of this study centered on photonic crystal
exciton polaritons, the most important photonic state is the
lowest energy state $\Delta_1$ 
located at the $\Delta$ point of the BZ. In this state, the
optical mode shows no phase shift at adjacent lattice sites, such
that oscillations of the electromagnetic field 
are in-phase at all microcavities composing the lattice. The intensity distribution 
of the $\Delta_1$ optical mode 
reveals 
periodic Bloch oscillations in the plane of QWs
[Fig.~\ref{fig2}(b), right panel]. As expected by the lattice
fill factor considerations ($FF{=}0.5$), this mode oscillates at
intermediate photon energy of 1.691 eV compared to the
vertical-cavity modes at the array pixel (1.687 eV at $FF{=}1$)
and at the grid (1.693 eV at $FF{=}0$). Correspondingly, the
energy of the ground state $\Delta_1$ can be modified within this
range by varying the lattice cell fill factor $FF$. It can be
then seen that a positive energy difference between a photon in
the $\Delta_1$ state and QW exciton is maintained at any $FF$ of
the lattice. [In Fig.~\ref{fig2}, the QW exciton of 1.682 eV
energy is located in between the energy scales shown in panels
(b) and (c).]

The coupled states of PhC photons and  QW excitons are analyzed
here using Jaynes–-Cummings model that takes into account the
coupling between a QW exciton and vacuum field oscillations in a
PhC mode. The top and bottom panels in Fig.~\ref{fig2} show,
respectively, the upper polariton (UP) and lower polariton (LP)
energy bands as well as the coordinate probability distribution
functions $|\psi(x,y)|^2$ for the upper and lower polariton
states of lowest energies (the states $\Delta_1^{UP}$ and
$\Delta_1^{LP}$), calculated using the Rabi coupling constant
$2\Omega_R=26$ meV.

It can be seen that due to a photonic constituent, the features
of periodic Bragg reflections in the plane of photonic crystal
lattice are transferred to exciton polaritons. Thus, the energy
dispersion curves of UP and LP (Figs.~\ref{fig2} (a) and (c),
left panels) show energy bands separated by forbidden gaps.
Respectively, the wave functions of 
exciton
polaritons exhibit periodic Bloch oscillations. In the case of
uniform photonic lattices, these periodic oscillations are
modulated with plane wave envelope functions propagating in the
plane of QWs.

As indicated by the modulation contrast of the $|\psi(x,y)|^2$
distributions,
due to contribution of excitons 
from flat QWs, the Bloch oscillations are less
pronounces in 
the UP and LP wave functions as compared to photons.
Furthermore, in Fig.~\ref{fig2}, 
the lower polariton states have higher excitonic content and they
are characterized by smoother energy bands and wave function
distributions, as compared to the upper polaritons. Nevertheless,
the features of
periodic Bloch oscillations 
are clearly visible in the LP wave functions 
and one can expect that the envelope functions of polaritons
can be tailored 
by using photonic crystal heterostructures.

\psection{Harmonic oscillator trap for exciton
polaritons}\label{SecHarmTrap}

\begin {figure}[tbp]
\center
\includegraphics {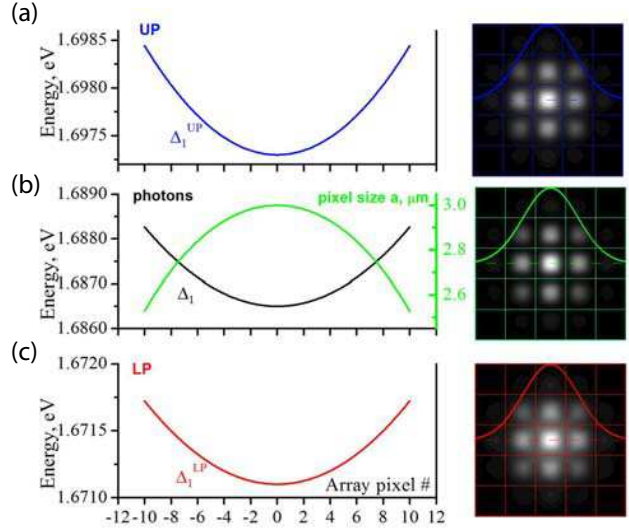}
\caption {
Parabolically graded band edges of lowest-energy bands (left panels) and 
ground-state probability densities $|\psi(x,y)|^2$ (right panels) 
of the upper polaritons (a), photons (b) and lower polaritons (c)
in the trap defined by tailored pixel-size variations [green
curve in (b), right axis] in a quasi-periodic 2D array of coupled
CdTe/CdMgTe microcavities ($\Lambda=3~\mu m$). The panels (a)-(c)
are ordered according the the energy scale. }\label{fig3}
\end {figure}

Trapping of exciton polaritons in a harmonic oscillator potential
is
accomplished 
here via 
their envelope wave functions. A photonic crystal heterostructure
trap
is defined 
by introducing tailored variations of the
cavity pixel size across the array structure.

A numerical analysis shows that for the low-contrast lattices
considered here and for the lattice cell fill factor FF in the
range of 0.1--1, the $\Delta_1$ photonic band edge energy
decreases monotonically with increasing cavity pixel size $a$.
Therefore, the pixel-size variations ${\propto} {-} (x^2+y^2)$
along the two lattice directions of an array
produce 
a parabolically graded 
shift ($\propto r^2$) of the photonic
band edge $\Delta_1$. 
Fig. \ref{fig3} (b)
shows variations of the 
pixel size with the
lattice position (right axis, green curve) and 
the resulting effective potential profile $U_P=\frac 12 \alpha_P
r^2 $  for photons (left axis, black curve).

When an 
exciton from a flat QW forms a coupled state with a photon in
the effective potential $U_P$, 
the
upper ($\Delta_1^{UP}$) 
and
lower ($\Delta_1^{LP}$) polariton bands 
exhibit parabolic variations of the band edge with the lattice
position as well.
These 
variations define the effective confining potentials 
$U_{UP}{=}\frac 12 \alpha_{UP} r^2 $ and $U_{LP}{=}\frac 12
\alpha_{LP} r^2 $ for the
upper and lower polaritons (Fig.~\ref{fig3}(a) and (c), respectively). 
In Fig.~\ref{fig3}, the effective force constants of
the trap 
are $\alpha_P{=}390$, $\alpha_{UP}{=}250$ and
$\alpha_{LP}{=}140~eV/cm^2$ for the photons, UP and LP states,
respectively.

The confined envelope functions of exciton polaritons 
in the trap are analyzed here using the effective mass 
approximation \cite{Boiko07B} with effective masses derived from
the energy dispersion curves in the vicinity of the $\Delta$
point of the BZ. The effective masses of a photon (in the plane
of QWs), UP and LP are, respectively,  $m_P=0.9\cdot
10^{-4}m_{e}$, $m_{UP}=1.4\cdot
10^{-4}m_{e}$ and $m_{LP}=2.4 \cdot 10^{-4}m_{e}$,  with $m_{e}$ being the free electron mass. 
The larger mass of lower polariton is due to the higher excitonic
content in LP states, in agreement with the features of periodic
Bloch
oscillations 
in probability density 
$|\psi(x,y)|^2$ (Fig.~\ref{fig2}).

The calculated 
trap force constants and effective masses assume that the
excitation energies $\hbar
\omega_{osc}=\sqrt{\alpha/m^*}$ of 
photonic, UP and LP oscillators 
are of
59, 38 and 21 $\mu eV$, respectively. These 
energies also
define a shift of 
corresponding ground oscillator states from
the bottom of the trap.

The wave functions of photons, UPs and LPs in the ground state of
the trap are indicated in Figs.\ref{fig3}(a)-(c) (right panels).
They exhibit periodic Bloch oscillations modulated with the
Gaussian envelope functions delocalized over several lattice
sites. The 
envelope functions for the  photons, UPs and LPs are almost
identical to each other. 
The apparent difference between the spatial distributions of
probability densities in Fig.~\ref{fig3} is caused by
dissimilar Bloch functions 
of photons and polaritons.

\psection{Results and discussion}\label{SecResDisc}

\begin {figure}[tbp]
\center
\includegraphics {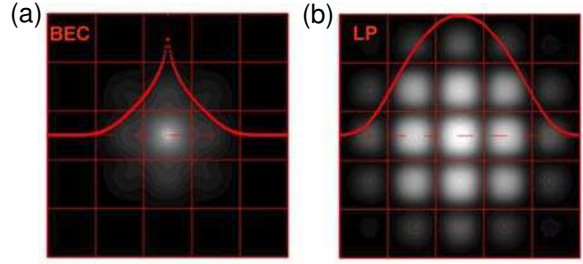}
\caption { Probability density distributions and envelope function
cross-sections for the LPs in the BEC fraction (a) and in the
ground oscillator state (b) plotted across a region of 5x5 lattice
sites of a quasi-periodic array of CdTe/CdMgTe microcavities (the
array pitch is $\Lambda=3~\mu m$). In both cases, a logarithmic
scale $\propto M \ln (1+ |\psi(x,y)|^2/p_{\text{th}})$ is used
with the probability density threshold of
$p_{\text{th}}=10^{-3}$, but different scale factors $M$ are
applied to the macroscopic wave function (a) and single-polariton
wave function (b). }\label {fig4}
\end {figure}

In previous sections, the interactions in the LP system have not
yet
been taken into account and the 
analysis of confined states in the trap has been carried out in
the effective mass approximation, using stationary Schr\"{o}dinger
equation. Introducing 
a term $\propto |\psi(x,y)|^2$, which accounts for small repulsive
interactions in the polariton system, 
one obtains 
a nonlinear Gross-Pitaevskii equation for lower polaritons. Its
solution at zero chemical potential represents a condensate
fraction of LPs. In Fig.\ref{fig3}(c), this state is located at
the bottom of the trap, at 21 $\mu eV$ below the ground oscillator
state. 

The 
macroscopic wave function 
of the lower polariton condensate 
is shown in Fig.
\ref{fig4}(a) (logarithmic scale).
The 
condensate fraction 
is localized to a single cavity pixel at the center of the trap.
The excitonic content of lower polaritons in this state is of 0.6.
For comparison, Fig.~\ref{fig4}(b) shows the LP wave function in
the ground oscillator state, plotted in the logarithmic scale as
well. The spatial distribution of polaritons in the condensate
fraction (a) is thus
significantly different from the one 
in the ground state of the trap (b). 
It even more
drastically differs 
from the spatial distribution of polaritons in non-condensate
fraction, which occupy the excited states in the trap within the
$k_B T$ energy range and are delocalized over a region of size
${\sim} \sqrt{2k_B T/\alpha_{LP}}$.
It should be stressed that this difference is due to 
the envelop function properties of lower polaritons in photonic
crystal lattices with small repulsive interactions in polariton
system.

For particular trap considered here,  the critical threshold
number of LPs needed to achieve the BEC phase transition can be
estimated from expression $N_c=1.8(k_B T)^2/ (\hbar
\omega_{osc})^2$ \cite{Negoita99}, yielding $Nc=4.4 \cdot 10^3$
polaritons in the trap at a temperature of 12 K. The
non-condensate fraction is delocalized over a region of 40 $\mu
m$ width (about 13 lattice sites), which makes it clearly
distinguishable from the condensate fraction at the center cavity
pixel of the trap (of $\sim 3 ~ \mu m$ width). The thermal
excitation energy $k_B T$ of about 1 meV significantly exceeds
the energy separation of quantized LP oscillator states in the
trap (21 $\mu eV$). Taking also the positive photon-exciton
energy detuning into account, one should expect that such trap
favors the thermalization of LPs and spontaneous BEC phase
transition in polariton system.

Finally note that
the periodic photonic 
lattice will prevent a localization of exciton polaritons due to
disorder effects in the quantum wells, observed in experiments on
k-space BEC.

\psection{Conclusion}

Photonic crystal heterostructures offer a powerful approach for
tailoring the envelope functions of exciton polariton modes
propagating in quasiperiodic arrays of microcavities. The use of
this concept is illustrated here on example of
photonic crystal heterostructure implementing 
harmonic
oscillator 
trap for exciton polaritons. This concept should stimulate
further development of novel applications of polariton-based
systems for control of exciton polariton propagation and
confinement, and in particular in delivering important evidence
that exciton-polaritons might undergo BEC phase transition.

\end{paper}

\end{document}